\documentclass[preprint,12pt]{elsarticle}

\usepackage{amssymb}
\usepackage{graphicx}
\usepackage{csquotes}
\usepackage{hyperref}

\journal{Journal of Magnetic Resonance}

\begin{document}

\begin{frontmatter}



\title{Optically stimulated electron paramagnetic resonance: simplicity, versatility, information content}


\author[solab]{Kozlov V.~O.}
\author[solab]{Fomin A.~A.}
\author[solab,phot]{Ryzhov I.~I.}
\author[solab,ssd]{Kozlov G.~G.}

\affiliation[solab]{organization={Spin Optics Laboratory, Faculty of Physics, St Petersburg State University},%Department and Organization
            addressline={Peterhof, Ul'yanovskaya ul., 1}, 
            city={Saint Petersburg},
            postcode={198504}, 
            %state={},
            country={Russia}}
\affiliation[phot]{organization={Photonics Department, Faculty of Physics, St Petersburg State University},%Department and Organization
            addressline={Peterhof, Ul'yanovskaya ul., 1}, 
            city={Saint Petersburg},
            postcode={198504}, 
            %state={},
            country={Russia}}
\affiliation[ssd]{organization={Solid State Department, Faculty of Physics, St Petersburg State University},%Department and Organization
            addressline={Peterhof, Ul'yanovskaya ul., 1}, 
            city={Saint Petersburg},
            postcode={198504}, 
            %state={},
            country={Russia}}

\begin{abstract}
A simple technique for observing optically stimulated electron paramagnetic resonance (OSEPR) is proposed and investigated. The versatility and information content of the described technique is demonstrated by the example of the OSEPR spectra of systems that are unpopular for this type of spectroscopy: a crystal with rare-earth ions Nd$^{3+}$ and a doped semiconductor GaAs. In addition, the OSEPR spectrum of atomic cesium is presented, in which an optical nonlinearity is observed that makes it possible to estimate the Rabi frequency for the relevant optical transition. The effects observed in the described experiments (switching of peaks to dips, light-induced splitting of the OSEPR lines, and the appearance of a spectral feature at the double-Larmor frequency) are interpreted using the model proposed in the theoretical part of the work. The suggested interpretation shows the possibility of using the described OSEPR technique to estimate not only `magnetic' parameters of the model Hamiltonian (g-factors, spin relaxation times), but also the Rabi frequencies characterizing optical transitions.
\end{abstract}

\begin{graphicalabstract}
\includegraphics{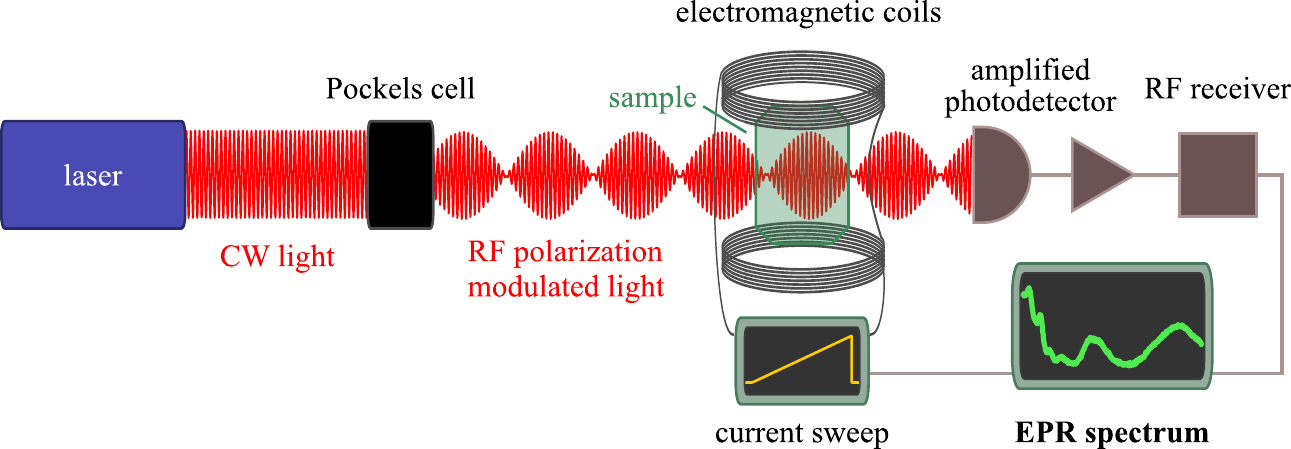}
\end{graphicalabstract}



\end{frontmatter}

%\definecolor{darkgreen}{RGB}{30,150,10}
%\definecolor{darkred}{RGB}{150,30,10}
%\definecolor{darkblue}{RGB}{30,10,150}
%\newcommand{\addIR}[1]{\textcolor{darkgreen}{#1}}
%\newcommand{\IR}[1]{\textcolor{darkgreen}{\textbf{IR:} \textit{#1}}}
%\newcommand{\addGK}[1]{\textcolor{darkred}{#1}}
%\newcommand{\GK}[1]{\textcolor{darkred}{\textbf{GK:} \textit{#1}}}
%\newcommand{\addVZ}[1]{\textcolor{darkblue}{#1}}
%\newcommand{\VZ}[1]{\textcolor{darkblue}{\textbf{VZ:} \textit{#1}}}

%-------------
%\oddsidemargin = -0.52cm %              отступ слева + 2,54 см
%\textwidth = 16.96cm %                  ширина содержимого (текста)
%\topmargin = 0cm %                      отступ сверху
%\textheight = 23.7cm %                  высота содержимого (текста)
%\voffset = -1.54cm %                    сдвиг страницы вверх
%\paperwidth = 21cm %                    ширина листа
%\paperheight = 29.7cm %                 высота листа
%\renewcommand{\baselinestretch}{1.1} %  Интерлиньяж (интервал между строками:))
%\parskip=6pt

\section*{Introduction}
 
Electron paramagnetic resonance (EPR) spectroscopy is one of the most important methods of experimental physics~\cite{zavoisky-epr45,altshuler-epr72,abragam-epr-ions70}. This type of spectroscopy is based on observation of the resonant precession of magnetic moments (spins) of the sample under study 
in an external magnetic field, whose characteristic frequencies (Larmor frequencies) carry information about the electronic and nuclear motion in the sample.
Among the various methods of EPR spectroscopy, those based on {\it optical} detection of EPR (ODEPR)\cite{optical_detection_magn_res_suter}, when dynamics of magnetic moments is observed as modulation of the polarization of a probe optical beam, 
 %passing through the sample,
 have become very popular in recent decades~\cite{zapasskii-evolution17en} (whilst the relationship between magneto-optics and EPR was pointed out as far back in 1951~\cite{kastler-optical-mr-study51}).
 
 ODEPR methods can be divided into the following three groups. The first group includes methods in which
 the excitation of the magnetic moments precession 
 %of the paramagnetic system 
 is performed by a high-frequency magnetic field, and the EPR is recorded using a probe laser beam~\cite{daniels-spin-latt-relax60,asawa-direct-opt-det62,aleksandrov-esr-tm-ion-fluorite-en77,bitter-optical-detection-rf49,yelin-nonlin-opt-double-dark-res03} or vice versa \cite{direct_detection_yoshihiro}. In the second group called below {\it optically stimulated} EPR (OSEPR), both excitation and registration of EPR are carried out in the optical channel~\cite{bell-bloom-opt-driven61,optical_excitation_novikov,alexandrov_dyn_eff_non_linear,pedreros_budker_pol_spin_prec,braggio_optical_manipulation,grujic-atomic-mr-induced10, budker-opt-magnetometry07,simin-all-opt-dc-nt-magnetomtry16}. 
  Both all-optical~\cite{alexandrov_dyn_eff_non_linear} and double radiooptical  magnetic resonance detection methods find its application in a wide area of optical magnetometry \cite{budker-opt-magnetometry07,simin-all-opt-dc-nt-magnetomtry16}, optical spin-related phenomena investigation and manipulation (e.g.~\cite{cavenett-odmr-invest-semicond81,shinar-odmr-oled12,negyedi-odmr-spectrometer17}).

 A special place is occupied by methods of the third group, the so-called spin noise spectroscopy (SNS), in which EPR is observed as a feature in the polarization noise spectrum of a probe beam at the Larmor frequency~\cite{aleksandrov-mr-fr-noise-en81,zapasskii-sns-review13,huebner-rise-of-sns14}. This method of EPR detection does not require any excitation and can, in some cases, be considered as non-perturbative \footnote{A non-turbative method for observing the magnetic resonance of protons, based on recording noise and using standard NMR equipment, is described in \cite{gueron-water-nmr-sn89}}.
 Before proceeding, it is advisable to briefly indicate advantages of the ODEPR methods inherent in the above groups.

 An important advantage of the ODEPR over conventional EPR technique, 
 is its 
 optical {\it spectral selectivity}
 which allows one to detect EPR spectra only of a group of particles selected by the probe beam tuned in resonance with their optical transition.

The second advantage of the ODEPR technique is its 
{\it spatial selectivity}, i.e.
 the possibility of getting EPR signal only from a certain small area of the sample, the position of which can be changed in
the process of the experiment. This property of the ODEPR 
takes advantage of 
sharply focused probe optical beams and can be used for tomography of paramagnetic samples~\cite{romer-spatially-resolved-sns09}. 
  The ODEPR technique can be used for remote observation of spin precession in the upper atmosphere \cite{pedreros_budker_pol_spin_prec}.

Each of the methods listed above
has its own specific features.
 Here, we want to emphasize an essential distinction of 
the methods of the second group (OSEPR), to which the proposed article is devoted: the methods of this group are fundamentally {\it optically nonlinear}, since the excitation of the magnetic moments precession occurs, in this case, due to polarization, intensity, or frequency modulation of the probe beam. 
On the one hand, the nonlinear nature of the OSEPR somewhat complicates interpretation of the corresponding experiments, and on the other hand, it makes it possible to obtain additional information about the system under study.
 
Turning to the purpose of this work, we note that the OSEPR technique is in most cases used to observe EPR in gases~\cite{bell-bloom-opt-driven61,grujic-atomic-mr-induced10,acosta-nonlin-mag-opt-rot06}: experiments with these systems go back to the classical work of Bell and Bloom~\cite{bell-bloom-opt-driven61}. In our opinion, the range of the OSEPR objects is currently unjustifiably narrow, and the theoretical models used allow for further development. Therefore, the main purpose of our publication is to show the simplicity and versatility of the OSEPR technique, to demonstrate its applicability to new classes of objects, and to show the information potential of this technique associated with the possible difference between the OSEPR spectra and conventional EPR spectra. 
 
 The work is organized as follows.
 The first section describes an experimental setup with a polarization (intensity)
 modulation of the probe beam and scanning of the magnetic field. The next section gives examples of the OSEPR spectra of samples 
 with significantly different `origins' of spins:
 %that differ significantly from each other: 
 an SrWO$_4$ crystal with an admixture of Nd, a Te-doped GaAs crystal, and vapors of atomic Cs. Brief comments on the obtained spectra are also given in this section.
 The third section describes one of the possible schemes for calculating OSEPR signals. In the frame of the linear intensity-polarimetric response 
 %is introduced and the 
 the concept of corresponding susceptibility (called {\it Stokes susceptibility}) is introduced. The fourth section presents and discusses the results of model calculations of the Stokes susceptibility. The possibility 
 %of alternative modes of measuring the OSEPR spectra is discussed, as well as the possibility 
 of improving the experimental setup are also addressed. 
 The results obtained in the work are summarized in the Conclusion.

 \section{Experimental setup }
 
 \begin{figure}
\centering
 \includegraphics[width=.8\columnwidth,clip]{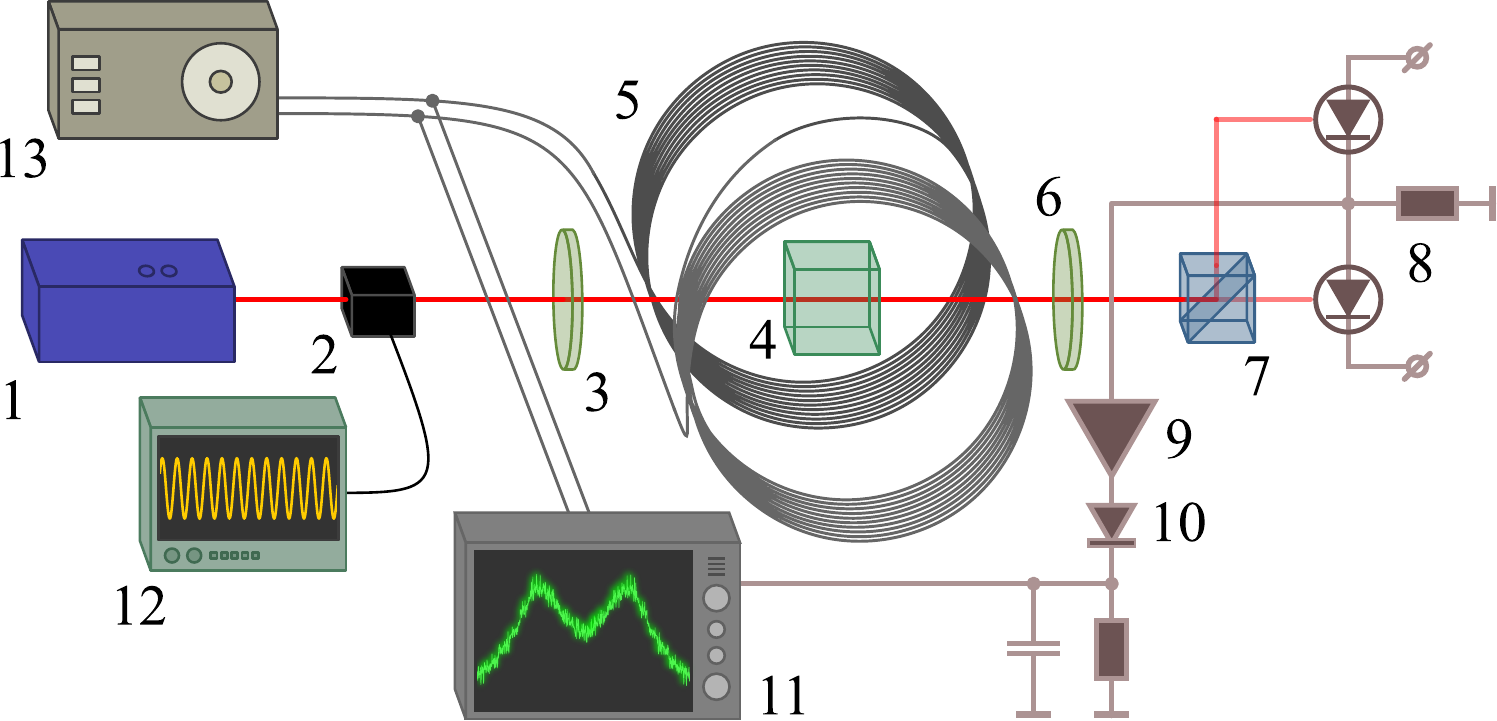}
 \caption{Scheme of the installation for the observation of optically stimulated EPR (OSEPR). Explanations in the text. }
 \label{fig1} 
 \end{figure} 
 
Schematic of our experimental setup for OSEPR observation is shown in Fig.~\ref{fig1}. A linearly polarized probe beam (approximately 2 mm in diameter) is generated by a tunable laser 1 and directed to a polarization modulator consisting of Pockels cell 2 and quarter-wave plate 3. Mutual orientation of the polarization plane of the laser beam, the electric field in the Pockels cell and the plate $\lambda/4$ are chosen so that when a voltage is applied to the Pockels cell, the polarization azimuth of the light at the output of the $\lambda/4$ plate rotates by an angle proportional to the cell voltage\footnote{In the above scheme, the azimuth of the polarization plane of the probe beam is modulated. This fragment of the scheme can be easily changed if modulation of the ellipticity or intensity of the probe beam is required.}. Next, the probe beam passes through sample 4 in a magnetic field transverse to the probe beam (Voigt geometry) created by electromagnet 5 and enters the polarimetric receiver, which responds to oscillations of the input beam polarization plane. The half-wave plate 6 serves for the initial balancing of the polarimetric receiver, which consists of polarization beamsplitter 7 and differential photodetector 8.
 From generator 12, a high-frequency (57 MHz) voltage is supplied to Pockels cell 2 with an amplitude of 1-2 $\%$ of the half-wave voltage of the cell ($\sim 1$ volt in our experiments). 
 %In this case, 
 The high-frequency signal of 57 MHz at the output of the differential photodetector 8 is intensified by the resonant amplifier 9 and detected by the receiver 10, whose output voltage is proportional to oscillation amplitude of the probe beam polarization plane at the exit of the sample. In the described version of the OSEPR, the dependence of this amplitude on the magnetic field 
 %in which the sample is located
 is recorded. This dependence---we call it the OSEPR spectrum---exhibits peculiarities when the modulation frequency (57 MHz) coincides with the magnetic field-dependent Larmor frequency of the magnetic moments. The OSEPR spectrum is observed on the screen of the oscilloscope 11, on the vertical plates of which a signal from the radio-frequency detector 10 is applied, while 
 the voltage on the horizontal plates is
 %on the horizontal plates - a voltage 
 proportional to the magnetic field. %taken from the power supply of the electromagnet 13. 
 In our setup, the magnetic field was varied 
 %could change over time
from 0 to approximately 100 gauss according to a triangular law at frequency of 20\ldots40 Hz. In this case we could observe the OSEPR spectrum directly on the screen of the oscilloscope 11 or (for essentially noisy signal) to quickly accumulate it using a digital computer system (not shown in Fig.~\ref{fig1}). The described OSEPR experimental setup is simple and provides a significant magnitude of the observed signals, which often do not require special processing and long accumulation times for their registration. 

Below, we give examples of the OSEPR spectra of various samples recorded at the setup Fig.~\ref{fig1} in the Voigt geometry. Unless otherwise noted, the probe beam was linearly polarized at an angle of 45$^o$ to the transverse magnetic field. The experiments with cesium vapor were carried out at a lower modulation frequency (20.5 MHz).
 
 \section{Examples of OSEPR spectra}
 
 In this section, we present several examples of OSEPR observations that demonstrate universality of this type of spectroscopy and its applicability to systems with essentially different `nature' of the magnetic moments (i.~e., the media under study are essentially different in physical properties and their noise responses are usually described in different model frameworks).
 
\subsection{Rare-earth ion in a dielectric crystal
%Nd$^{3+}$ ions in SrWO$_4$ crystal
}\label{RE}
 \begin{figure}
 \centering
 \includegraphics[width=.7\columnwidth,clip]{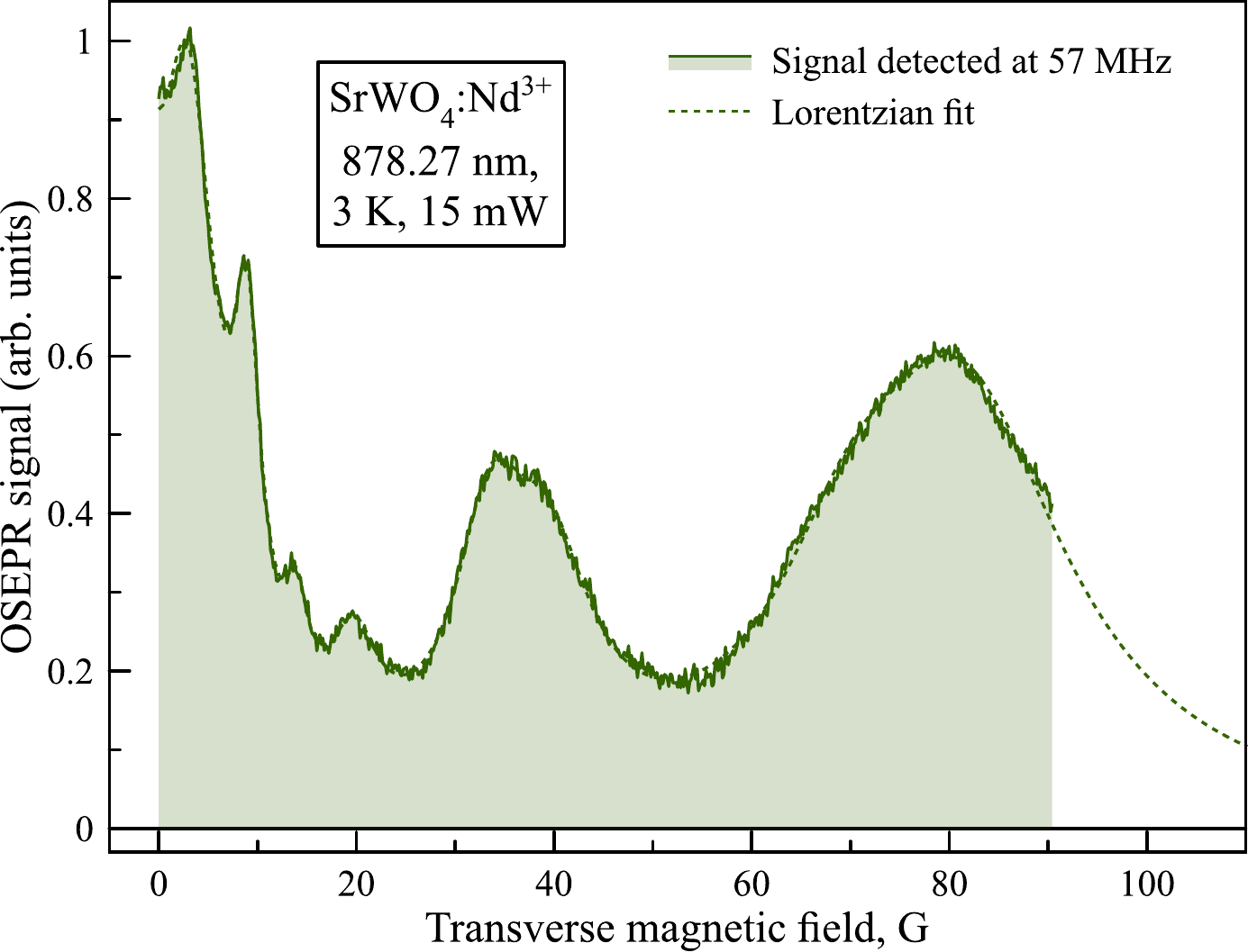}
 \caption{
 The OSEPR spectrum of the SrWO$_4$:Nd$^{3+}$ crystal recorded with a probe beam polarization plane azimuth oscillation at 57 MHz. The sample length was approximately 5 mm. The probe light characteristics are provided on the figure, the focusing was done with 75 mm lens and 2 mm diameter of a beam. The smooth curve is the spectrum fit with 7 Lorentzian contours.
 }
 \label{fig2} 
 \end{figure} 
Crystals with rare-earth impurities have been actively studied by traditional EPR spectroscopy (see, e.\,g.,~\cite{sukhanov-crystal-env-impurity18,gafurov-epr-study-some-re-ions03,parekh-magn-dynamics-re-gd12}),  however, we are not aware of OSEPR observations in such systems. Figure \ref{fig2} shows the typical OSEPR spectrum of Nd$^{3+}$ ions (1~mol.\% concentration) in the SrWO$_4$ crystal registered on our setup at liquid helium temperatures\footnote{The SrWO$_4$ crystal doped with Nd$^{3+}$ was kindly provided to us by L. I. Ivleva (A. M. Prokhorov Institute of General Physics, Russian Academy of Sciences).}.
 Since the Nd$^{3+}$ ion in the crystal matrix is in the electric field of the crystal atoms surrounding it, its optical and EPR spectra differ significantly from those of the isolated ion. The variety of possible environments determines
 appearance of various groups of paramagnetic centers associated with the Nd$^{3+}$ ion. This leads to the complication of the corresponding OSEPR spectrum.
 As can be seen from Fig.~\ref{fig2}, the spectrum consists of a large number of lines (fitting gives 7 lines) with significantly different widths and effective g-factors differing by more than an order of magnitude.
 Panoramic recording of such a spectrum in one experiment using a conventional EPR spectrometer,
 operating at frequencies $\sim$10\ldots30 GHz,
 requires a change in the magnetic field $\sim$2.5~T, which is not always possible, not to mention field scanning in such a range with a frequency of 20\ldots40 Hz, as is the case in our experiments.
 Passing to lower frequencies ($\sim$60 MHz) in the traditional EPR technique is associated with a decrease in sensitivity and is usually accompanied by an increase in the volume of the sample under study to $\sim$1 cm$^3$ (the volume of the oscillatory circuit coil at 60 MHz). In the described version of the OSEPR, the volume of the sample, from which the signal was recorded, was determined by the volume of the probe beam in the sample and was estimated\footnote{Given as Rayleigh length multiplied by focused beam waist.} at 10$^{-6}$ cm$^3$.
 
We note that the above-mentioned spectral selectivity of the OSEPR in the optical channel can be very useful in deciphering the EPR spectra recorded by the traditional method.
 We had at our disposal the EPR spectrum of our sample SrWO$_4$:Nd$^{3+}$, recorded on a conventional
 EPR spectrometer, which demonstrated a much larger number of features than the OSEPR spectrum in Fig. \ref{fig2}. The purpose of this publication is not to decipher the EPR spectrum of SrWO$_4$:Nd$^{3+}$, however, it is clear that a comparison of the spectra of traditional EPR and OSEPR will help to identify the paramagnetic centers of the sample under study, in particular, thanks to additional optical selectivity of transitions in OSEPR. 
 
\subsection{A doped semiconductor 
%Impurity GaAs
}\label{GaAs}
 
\begin{figure}
\centering
\includegraphics[width=0.6\columnwidth]{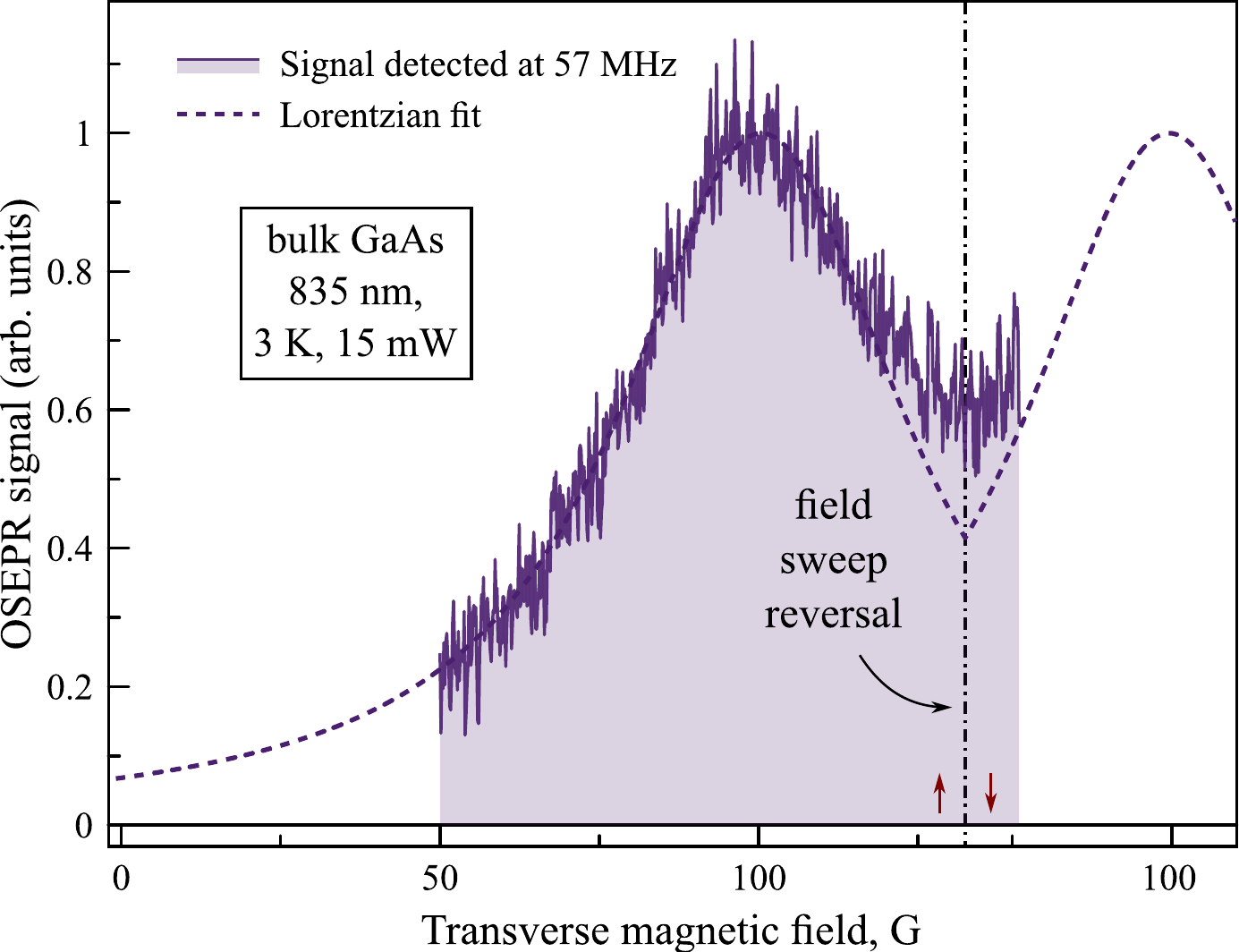}
\caption{OSEPR spectrum of a GaAs crystal doped with tellurium (electron density $\sim$2$\cdot 10 ^{16}$ cm$^{-3}$, sample thickness $\sim$2 mm). The sweep was performed in the presence of a permanent magnet, which shifted the magnetic field range. Geometry of the probe beam was the same as in Fig.~\ref{fig2} The smooth dashed line is a fit with a Lorentzian in the range up to\, $\sim$132 G, where magnetic field starts to decrease. The slight discrepancy of the fit is ascribed to the imperfection of the sweep triangular law (especially at the turning point).}
\label{fig3} 
\end{figure} 
As a second example of OSEPR, we consider the application of this technique to the study of a tellurium (Te) doped GaAs semiconductor crystal (electron concentration $\sim$2$\cdot 10 ^{16}$ cm$^{-3}$).
In contrast to the Nd$^{3+}$ ionic system considered above, in this case the polarization-optical response is generated by the electron gas arising, in the conduction band, due to doping\footnote{The EPR spectrum of this sample recorded using the optical spin noise spectroscopy technique is given in~\cite{glasenapp-polarimetric-sens13}.}.
The typical OSEPR spectrum of this GaAs:Te sample recorded at our setup at low temperature is shown in Fig.~\ref{fig3}.
 The spectrum in Fig.~\ref{fig3} qualitatively coincides with the noise spectrum~\cite{glasenapp-polarimetric-sens13} and represents a single rather wide (in frequency units $\sim$20 mHz) line corresponding to the $g$-factor $g\approx 0.4$.
 Note that the spectrum in Fig.~\ref{fig3} was observed at a probe beam wavelength below the band gap of GaAs, where the sample was transparent. As the probe beam wavelength approaches the fundamental absorption edge of GaAs, the signal significantly increases (as in~\cite{glasenapp-polarimetric-sens13}), making it possible to observe it on the oscilloscope screen without accumulation.
 
 Spin state manipulation by means of optical pulses in semiconductor systems like GaAs:Te is well known and actively used~\cite{kikkawa-resonant-spin-amp98, dzhioev-spin-relax02, chen-opt-excit-spins-qw10}.
   Moreover, the paper \cite{all_optical_magnetic_res_awschalom} describes a method for observing nuclear magnetic resonance under optical excitation. 
 However, the corresponding experiments are performed, as a rule, using the pulsed pump-probe technique~\cite{awschalom-dyn-spin85} and differ from those described above both in the equipment used and methodologically.

\subsection{Atomic vapor}\label{Cs}
 
\begin{figure}
\centering
\includegraphics[width=.7\columnwidth,clip]{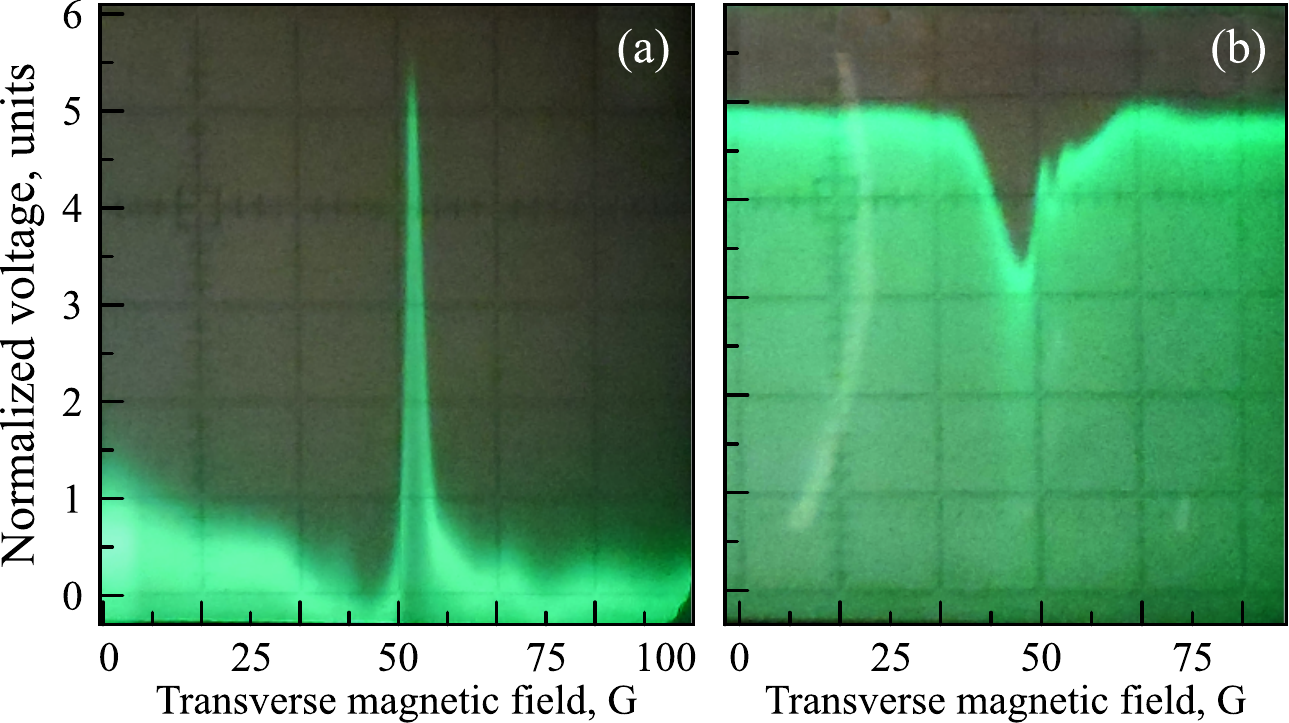}
\caption{OSEPR spectrum of Cs atomic vapors at room temperature.
The modulation frequency is 20.5 mHz, the magnetic field range is 0\ldots100 gauss.
(a) The frequency of the probe beam corresponds to the slope of the short-wave component of the D$2$ line (transition from $F=3$), the probe beam power is 1.6 mW. (b) The frequency of the probe beam corresponds to the center of the long-wavelength component of the 
D$2$-line (transition from $F=4$), probe beam power 6 mW. One can clearly see the splitting of the OSEPR line, which makes it possible to estimate the Rabi frequency of the optical transition $\sqrt{|a_x|^2+|a_z|^2}\sim 5$ MHz. }
\label{fig4} 
\end{figure} 
Finally, let us describe the OSEPR of cesium vapor, a classical object of atomic spectroscopy.
 The sample ((4) in Fig.~\ref{fig1}) in this case was a cell $\sim$ 2 cm long with a small amount of metallic cesium in a special process. The cell was at room temperature, which corresponded to the concentration of saturated cesium vapor in the cell $\sim$ 10$^{12}$ cm$^{-3}$. The experiments were carried out at a wavelength of the probe beam corresponding to the D$2$ line (852 nm).
 The OSEPR spectrum (Fig.~\ref{fig4}) in this case is observed directly at the output of the polarimetric receiver with a signal-to-noise ratio of $\sim$20~---
 resonant amplifier 9 and detector 10 (see Fig.~\ref{fig1}) were removed. Recall that in our experiments, we observed a magnetic-field-dependent response of the sample to the modulation of the azimuth of the polarization plane of the probe beam. This response represented an increase (or decrease) of the above modulation when its frequency coincided with that of Larmor precession of the magnetic moments. In such an arrangement of the experiment, the polarimetric receiver always registered a background signal independent of the magnetic field (also observed in the absence of the sample). For solid-state samples, the OSEPR signal was a few percent of the background signal. The OSEPR signal of Cs vapor was more than 10\% of the background signal, which could be removed by shifting the electron beam of the analog oscilloscope (Fig.~\ref{fig4}a).
 
 For the gaseous system under consideration, the character of the OSEPR spectrum essentially depends on intensity, spectral position, and polarization of the probe beam. Since the purpose of this article is to demonstrate the methodological possibilities of the OSEPR, we will restrict ourselves here to brief description of the effects of inversion, broadening, and splitting of the OSEPR spectra.
 
 Figure \ref{fig4}b shows the OSEPR spectrum of Cs vapor when a relatively powerful probe beam is tuned to the center of the long-wavelength component of the D$2$ line. As can be seen from Fig.~\ref{fig4}b, the character of the spectrum changes significantly - the resonant peak in Fig.~\ref{fig4}a is replaced by a dip, and a significant broadening and splitting of the OSEPR line appears. Fig.~\ref{fig4}b also gives an idea of the magnitude of the background signal discussed above.
 The theoretical consideration given in the next section explain %shows the possibility of 
 the observed splitting of the OSEPR spectrum in the optical field of the probe beam. Its value can be estimated using the spectrum in Fig.~\ref{fig4}b ($\sim$5 MHz), which, in turn, allows us to estimate the Rabi frequency of the optical transition of Cs atoms in our experiment.
 
 In conclusion of this subsection, we note that it was not difficult to observe a signal at the second harmonic of the Larmor frequency on our experimental setup~--- the possibility of such an effect is described, for example, in~\cite{dyakonov-theory-res-scattering-light-gas65,grujic-atomic-mr-induced10}

 \section{Theoretical consideration}\label{theory}
 
The excitation of the precession of magnetic moments with the help of light was considered theoretically, for example, in
\cite{grujic-atomic-mr-induced10, mitsui-sns-atomic00, yabuzaki-new-type-spectroscopy90, camparo-obs-rabi-spectrum88, camparo-rabi-off-resonant98, camparo-convers-phase-noise99, godone-rabi-lambda-scheme02, dyakonov-theory-res-scattering-light-gas65,kozlov-intens-related-suscept-asns14}. Our model is oriented to the above-described OSEPR recording technique and allows, in principle, to solve the problem of calculating the polarization-intensity response of an atomic system to an arbitrary (frequency, polarization, intensity) modulation of the probe beam. Suggested model:

1) demonstrates, in fact, the possibility of OSEPR,

2) shows the commensurability of the OSEPR and background signals for the realistic parameters of the sample and the probe beam,

3) describes the optical splitting of the OSEPR lines and the possibility of changing the character of the OSEPR spectrum (replacing its peak by the dip),

4) shows the possibility of the OSEPR at the second harmonic of the Larmor frequency, and demonstrates a number of other properties of this signal.

In addition, despite the fact that OSEPR is an optically nonlinear phenomenon, in the formalism developed below, it turns out to be possible to introduce a susceptibility that describes the linear response of the system to a small polarization-intensity modulation of the probe beam.

Before proceeding to the description of the model, we note that a comprehensive analysis of the nonlinear propagation of an electromagnetic wave through a resonant medium
 is rather complicated, so we used the following approximations:

1. When describing atomic dynamics, we take into account only two spin-orbital multiplets---the ground and the excited ones---the 
 total angular momentums
%total moments 
of which $F$ we consider to be the same.

2. Probing is performed by a quasi-monochromatic optical beam, the frequency of which is detuned from
 that of the transition between the ground and excited multiplets by $\Delta$.
 
3. The relative change in the amplitude of the probe beam when passing through the layer of the resonant medium is small (thin-layer approximation).

Let's move on to the problem statement.
Let us introduce a coordinate system with the $z$ axis parallel to the magnetic field, the $y$ axis parallel to the probe beam propagation direction, and the $x$ axis orthogonal to the $z$ and $y$ axes. In this system, the probe beam at the input and output of the sample can be characterized by the Jones vectors $\bf a$ and $\bf u$, respectively:
 \begin{equation}
 {\bf a}=\left (\matrix {a_x\cr a_z} \right ),\hskip5mm
 {\bf u}=\left (\matrix {u_x\cr u_z} \right )
 \label{1}
 \end{equation} 
The projections of the probe beam field here will be considered to have the dimension of frequency. For example, $a_i\equiv A_i d/2\hbar,i=x,z$,
where $A_i$ are the projections of electric
field of the probe beam at the entrance to the sample, and $d$ is the dipole moment of the optical transition between
the ground and excited multiplets of our atomic system. The quantities $a_i$ determined in this way correspond to the Rabi frequency of the optical transition between the ground and excited multiplets.
 It can be shown that, in the used thin-layer approximation, the Jones vectors $\bf a$ and $\bf u$ are related by the following equations:
 \begin{equation}
\cases{u_{x,z}=a_{x,z}- \imath\omega_s \hskip1mm\hbox{Sp }\sigma D_{x,z}^- \cr
\imath \dot \sigma =[W,\sigma]+ R\sigma },\hskip5mm\hbox{ where }
\hskip5mm D_{x,z}^-\equiv \left (\matrix{0 & 0 \cr J_{x,z} & 0}\right )
\hskip5mm \omega_s\equiv {2\pi N kl d^2\over \hbar}
\label{2}
\end{equation} 
 \begin{equation}
W=W({\bf a})\equiv\Delta\left (\matrix {I & 0\cr 0 &0}\right )+ 
\left (\matrix {\omega_{L2}J_z & 0\cr 0 &\omega_{L1}J_z}\right )+ 
\left (\matrix{0 & J_x a_x^\ast+J_za_z^\ast\cr J_x a_x+J_z a_z & 0}\right )
\label{3}
\end{equation}
Here, $W$ and $\sigma$ are, respectively, the Hamiltonian of the atomic system and its density matrix in the interaction representation;
$N$ is the concentration of atoms; $k=2\pi/\lambda$ is the wavenumber of the probe beam;
$l$ is the length of the probed atomic system;
$\omega_{Li}=g_{Li}\mu_B B/\hbar, i=1,2$ --
 magnetic-field-dependent Larmor frequencies of the ground and excited multiplets (here, $g_{Li}$ is the Lande factor of the $i$-th multiplet, $\mu_B$ is the Bohr magneton, and
 $B$ -- is the external magnetic field); $\Delta$ is the optical detuning;
 $J_{x,z}$ -- $(2F+1)\times(2F+1)$ are the matrices of the $x$- and $z$- projections of angular momentum $F$; $R$ is the relaxation operator that phenomenologically takes into account damping
 in optical and magnetically split transitions.

If the Jones vector at the output of the sample is known, then the signal of the polarimetric receiver (operating in an arbitrary mode) can be calculated using standard methods of polarization optics. Thus, our task is to calculate the modulation of this vector given the modulation of the Jones vector at the input of the sample. If this modulation is small, then this problem can be considered within the framework of the linear response theory as follows. Let us assume that a constant probe beam with the Jones vector $\bf a$ acts on the atomic system for a long time. Let us denote by $\sigma_i$ and $\bf u$, respectively, the steady-state density matrix and the Jones vector of the probe beam at the sample exit.
 The quantities $\sigma_i$ and $\bf u$ defined in this way can be found by solving system (\ref{2}) with $\dot\sigma=0$. Let us now provide the input Jones vector with a small increment (jump) $\bf a\rightarrow a+\delta a$. In this case, the output Jones vector will also experience the increment ${\bf u}\rightarrow {\bf u}+\delta {\bf u}(t)$. The vector ${\bf\delta u}(t)$ describes the transient process in the atomic system at the jump $\delta \bf a$ of the input Jones vector and can be found by solving system (\ref{2}),(\ref{3}) for $\bf a\rightarrow a+\delta a$ with the initial condition $\sigma(0)=\sigma_i$. For example, if in an OSEPR experiment a sample is probed by a linearly polarized beam with a small modulation of its azimuth, then ${\bf a}=a\left ( \matrix {1\cr 1}\right )$ and $\delta{\bf a }=\delta a \Theta(t)\left ( \matrix {1\cr -1}\right ), \delta a\ll a$.
 With ${\bf \delta u}(t)$ known, one can find the output signal of the used polarimetric receiver $\delta s(t)$, which is the polarimetric response of the atomic system to the jump $\bf \delta a$ of the Jones vector of the input beam.

With the input beam harmonic modulation $\delta a\rightarrow \delta a_\nu e^{-\imath\nu t}$ used in our OSEPR experiments,
 the observed polarimetric response $\delta s\rightarrow \delta s_\nu e^{-\imath\nu t}$
 can be described by a susceptibility $\chi(\nu)$ such that $\delta s_\nu=\chi(\nu)\delta a_\nu$.
It is known, from the linear response theory, that this susceptibility is related to the response to
 jump $\delta s(t)$ by the relation
 \begin{equation}
\chi(\nu)=-\imath\nu\int _0^\infty \delta s(t)\hskip1mm e^{-\varepsilon t} e^{\imath \nu t} \hskip1mm dt, \hskip15mm \varepsilon\rightarrow +0 
\label{4}
\end{equation}
Susceptibility (\ref{4}) depends on all parameters included in the system (\ref{2}),(\ref{3}).
 In the OSEPR experiments described above, in fact, the dependence of $|\chi (\nu)|$ on the applied magnetic field is observed
at a fixed modulation frequency $\nu$. It is natural to call this type of susceptibility the {\it Stokes susceptibility}~\cite{grujic-atomic-mr-induced10,kozlov-intens-related-suscept-asns14}.

\section{Discussion}

From a formal point of view, the problem of finding the response $\delta s(t)$ of our atomic system to the jump of the Jones vector $\bf\delta a$ of the input beam and calculating the susceptibility $\chi(\nu)$ (\ref{4}) is reduced to solving the equation for the density matrix (the last equation of the system (\ref{2})).
 This is a linear equation and its solution can be obtained by diagonalizing the corresponding matrix,
 whose eigenvalues corrrespond to the spectral features (resonances) of the desired susceptibility $\chi(\nu)$. In the absence of relaxation ($R=0$), these features would correspond to the transition frequencies of the Hamiltonian (\ref{3}). Accounting for relaxation leads to the appearance of complex
 eigenvalues, which corresponds to the broadening of $\chi (\nu)$ susceptibility resonances. We do not dwell here on the calculation procedure, which is close to that described in~\cite{kozlov-intens-related-suscept-asns14}.
 
\begin{figure}
 \centering
 \includegraphics[width=.7\columnwidth,clip]{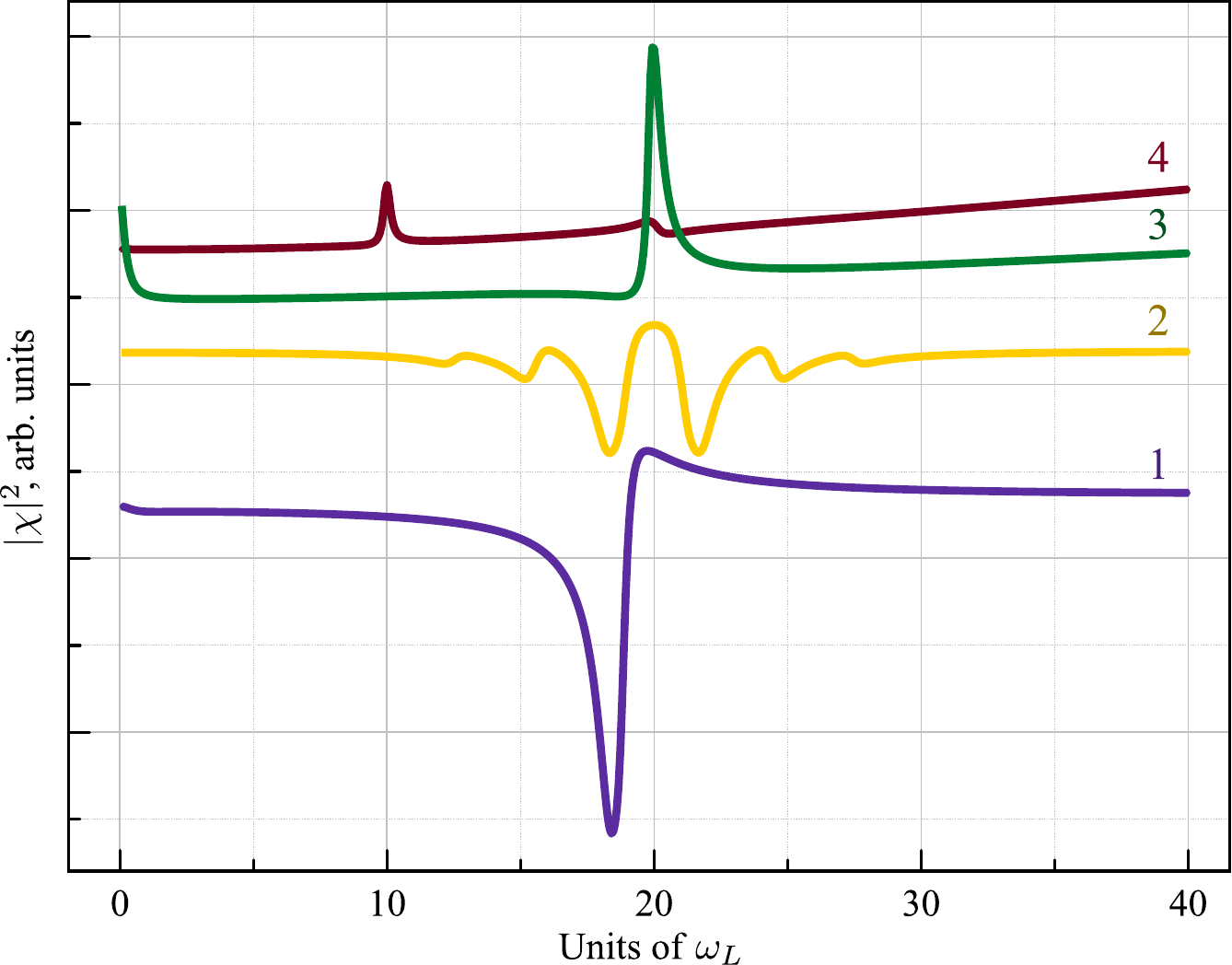}
 \caption{Theoretical dependences of $|\chi|^2$ on the magnetic field in units of the Larmor frequency $\omega_L$ (we put $\omega_{L1}=\omega_{L2}\equiv \omega_L$) for a given probe beam modulation frequency $\nu=20$. Curves 1 (total moment $F=1$) and 2 ($F=3$) were obtained with a relatively long decay time in the optical transition and probe beam polarization parallel to the magnetic field. In this case, the spectrum
 has the character of minima and at a large total moment of multiplets ($F=3$, curve 2) reveals a splitting estimated as $\sim \sqrt{|a_x|^2+|a_z|^2}$, the formula (\ref{3}). As the decay time decreases by several times, the character of the spectrum changes (the minimum is replaced by a maximum, curve 3), and when the probe beam is polarized at 45$^o$ to the magnetic field, a maximum appears at the second harmonic of the Larmor frequency (curve 4). }
 \label{fig5} 
 \end{figure}

Examples of the OSEPR spectra obtained using the model described above are shown in Fig.~\ref{fig5} along with brief explanations.
 The probe beam amplitude (more precisely, the corresponding Rabi frequency), for the presented model spectra, was
 $\sqrt{|a_x|^2+|a_z|^2}\approx 2\ldots4$. In addition, Fig.~\ref{fig5} gives an idea of the relative magnitude of the background signal discussed above.
 As can be seen from Fig.~\ref{fig5}, within the framework of our model, it is possible to reproduce the main qualitative properties of the OSEPR spectra in Fig.~\ref{fig4} listed at the beginning of the previous section. Comparison (at a qualitative level) of the calculated spectra with the experimental ones makes it possible to estimate the model parameters (\ref{2}), (\ref{3}). We note the information content of the described variant of OSEPR, which makes it possible to estimate not only the `magnetic' parameters determined in standard experiments on EPR and ODEPR (g-factors, spin relaxation times), but also the Rabi frequency of the optical transition (see Fig.~\ref{fig4}b ). 

The dynamics of the considered model system is described by the Hamiltonian (\ref{3}), whose eigenvalues (and the corresponding transition frequencies) are determined not only by the magnetic Zeeman contribution (the second term in Eq.(\ref{3})), but also by the third term, which depends on the amplitude of the probe
optical beam. Just as modulation of the Zeeman term leads to classical EPR, modulation of the third term (\ref{3}) leads to transitions between the states of the Hamiltonian (\ref{3}), which
are observed during OSEPR.
 Therefore, the OSEPR spectrum may differ from the conventional EPR spectrum, showing splittings and shifts associated with optical pumping by the probe beam (Fig.~\ref{fig4}b and Fig.~\ref{fig5}, curve 2).
 
We do not give here a rather cumbersome description of computational programs --- at this stage it seems redundant.
The considered simplest two-multiplet model cannot claim to be a quantitative description of the 
OSEPR spectra, for example, of Cs vapors (Fig.~\ref{fig4}), since, in this case, each component of the D$2$ line is formed by transitions of at least three multiplets. In addition, the type of relaxation
operator and allowance for Doppler broadening require special consideration, which is beyond the scope of this article. Also, 
 a special consideration is required by the OSEPR in a semiconductor system (Section 2.2).

%Insignificant changes in the experimental setup
  Relatively minor changes in the experimental setup  
(Fig.~\ref{fig1}) make it possible to observe OSEPR with arbitrary polarization-intensity modulation of the probe beam and with an arbitrary receiver mode\footnote{The receiver can operate in the Faraday rotation, ellipticity, or intensity registration mode}. For example, it is possible to modulate the intensity of the probe beam, while observing oscillations in the azimuth of the polarization plane, etc. The OSEPR measurements in this `cross' mode can provide additional information about the system under study.
 The calculation of such OSEPR spectra (i.e., the corresponding `crossed' Stokes susceptibility (\ref{4})) can be performed according to the scheme outlined above.

Finally, we note that the described setup and measurement procedure can be improved.
For example, it is possible to suppress the background signal by branching the probe beam in front of the sample and directing it to a separate receiver, the signal of which is subtracted from the signal of the main receiver. The use of digital lock-in amplification of the polarimetric signal can also increase the sensitivity of the OSEPR measurements. We plan to implement these and other modifications in the future.

\section{Conclusion}
 
We propose a simple yet universally applicable setup for observing optically stimulated electron paramagnetic resonance (OSEPR), also called optically driven spin precession. By means of the presented setup the versatility and information content of the implemented OSEPR realization are demonstrated. The experimental OSEPR spectra of carriers with essentially different `nature' of paramagnetism are presented: a paramagnetic centers in rare-earth-doped dielectric crystals, an electron gas in donor-doped semiconductor, and vapor of alkali metal. These OSEPR spectra were obtained by sweeping the magnetic field at a frequency of $\sim$20\ldots40 Hz without long-term accumulation.
We developed a theoretical model introducing the concept of the Stokes susceptibility. In this framework the description of the experimentally observed OSEPR spectra qualitative features is presented.

\section{Acknowledgments}
IIR and AAF acknowledge the financial support from Russian Science Foundation (Grant No.~21-72-10021) of experimental work on RE doped dielectrics and cesium vapor (presented in Sections~\ref{RE} and~\ref{Cs}, respectively). {VOK highly appreciates the Ministry of Science and Higher Education of the Russian Federation Megagrant №075-15-2022-1112 for funding the optical experimental studies of section~\ref{GaAs}). GGK acknowledges the SPbSU Grant No.~94030557 for the support of theoretical research (Section~\ref{theory})}. The work was fulfilled using equipment of the Resource Center ``Nanophotonics'' of the SPbSU Science Park.

\label{}



%\bibliographystyle{elsarticle-num} 
%\bibliography{osepr.bib}


%\begin{thebibliography}{00}


%\bibitem{}

%\end{thebibliography}

\end{document}